\documentclass[conference]{IEEEtran}
\IEEEoverridecommandlockouts
\usepackage{cite}
\usepackage{amsmath,amssymb,amsfonts}
\usepackage{algorithmic}
\usepackage{graphicx}
\usepackage{textcomp}

\usepackage{multirow}
\usepackage[colorlinks,citecolor=black,urlcolor=black,linkcolor=black,bookmarks=false,hypertexnames=true]{hyperref} 

\usepackage{array}
\newcolumntype{P}[1]{>{\centering\arraybackslash}p{#1}}
\newcolumntype{M}[1]{>{\centering\arraybackslash}m{#1}}

\def\BibTeX{{\rm B\kern-.05em{\sc i\kern-.025em b}\kern-.08em
    T\kern-.1667em\lower.7ex\hbox{E}\kern-.125emX}}
\begin{document}

\title{A Study of Consumers Cognitive Load in eCommerce Websites using Eye-tracking Technology\\
%{\footnotesize \textsuperscript{}} %subtitle if required
%\thanks{} %thanks notes if required
}
\author{\IEEEauthorblockN{1\textsuperscript{st} Shojibur Rahman}
\IEEEauthorblockA{\textit{Faculty of Science and Technology (FST)} \\
\textit{American International University-Bangladesh (AIUB)}\\
Dhaka, Bangladesh \\
shojiburman@gmail.com}
\and
\IEEEauthorblockN{2\textsuperscript{nd} Ahmed Alif Swopno}
\IEEEauthorblockA{\textit{Faculty of Science and Technology (FST)} \\
\textit{American International University-Bangladesh (AIUB)}\\
Dhaka, Bangladesh \\
ahmedalif1010@gmail.com}
\and
\IEEEauthorblockN{3\textsuperscript{rd} Nayeem Ahmed}
\IEEEauthorblockA{\textit{Faculty of Science and Technology (FST)} \\
\textit{American International University-Bangladesh (AIUB)}\\
Dhaka, Bangladesh \\
nayeemr.45@gmail.com}
\and

\IEEEauthorblockN{4\textsuperscript{th} Ashik Ahmed Fahim}
\IEEEauthorblockA{\textit{Faculty of Science and Technology (FST)} \\
\textit{American International University-Bangladesh (AIUB)}\\
Dhaka, Bangladesh \\
ashikahmed96@gmail.com}

\and

\IEEEauthorblockN{5\textsuperscript{th} Prof. Dr. Tabin Hasan}
\IEEEauthorblockA{\textit{Faculty of Science and Technology (FST)} \\
\textit{American International University-Bangladesh (AIUB)}\\
Dhaka, Bangladesh \\
tabin@aiub.edu}

}

\maketitle

\begin{abstract}
The aesthetics of e-commerce websites have a big influence on purchasing decisions and customers’ satisfaction. Webpage complexity and high cognitive load are responsible for causing an unpleasant experience while shopping online. This research empirically inspects a correlation between users’ cognitive load and product pricing, where price plays a vital role in causing web complexity. Therefore, we have experimented on 48 random individuals using eye-tracking technology to observe the eye movement calibration on some reputed e-commerce websites. We measured the cognitive load extracted from users’ datasets by analysing fixation count, saccades, fixation duration, and task completion time. Our study induces new findings on website complexity which varies on the similar product but different price ranges. This research also demonstrates a strong connection between customer perception and visual complexity while making online purchases. In addition, these findings will assist the developers and business analysts to improve consumers' shopping experience in e-commerce websites.

\end{abstract}


\section{Introduction}
    In many affluent and developing countries, the e-commerce industry continues to grow every year. It is expected that by the year 2025, the market dominance worldwide of e-commerce platforms will rise to 11 trillion \cite{b1}. Although the global economy suffered during the COVID-19 situation, the traditional market moved rapidly online to sustain itself in the business. Besides, strict quarantine and restrictions force consumers to become more internet centred \cite{b37}. E-commerce transactions increased rapidly during the pandemic in the US alone, from 10\% to 35\% \cite{b38}. The transparency and flexibility of e-commerce platforms have become more appealing to the vast majority in this global crisis.
    
    Since the mid-90s, the advancement decade of the e-commerce culture, it has always been a challenge to ensure the quality of the e-commerce website so that both company and customer can be benefited. The strength of the economy is determined by consumer choices \cite{b40}. Every e-commerce website's ultimate purpose is to influence users' buying intentions in order to sell items. A good design can make an e-commerce website succeed by encouraging desirable user behaviours that enhance the probability of a purchase. Tractinsky and Lowengart \cite{b39} proved that consumer purchasing behaviour depends on the aesthetics of a website. To elevate the e-commerce website's efficiency, we have studied the cognitive load by taking into account users’ visual perception.
    
    Visual perception means what the brain feels after observing the surroundings, and cognitive response means what the brain understands and reacts to it. As many varieties of people live worldwide, every person may have their unique perception, perspective, and views. Therefore, Some people prefer particular aesthetics, which others may despise. Visual perception is essential in online shopping because it allows users to understand better what attracts customers' attention online, how they absorb stimuli, and how they make purchasing decisions \cite{b41}. People require much time to perform a visual task on a complex website. When a website has a high cognitive load, it is classified as complicated. If consumers experience too many complexities while surfing product displaying pages, it might affect their mental models of purchasing behaviour \cite{b2}. On the other hand, low visual complexity means minimal cognitive load, which tends to have higher usability, accessibility and give higher user satisfaction \cite{b4}.
    
    However, Investigating cognitive load is one of the most challenging aspects because of the unpredictability of human intention. Substantial Human-Computer Interaction (HCI) research discusses the relationship between web page complexity and cognitive load without a decisive ending. A high cognitive load is always responsible for unsatisfactory interactions \cite{b45}. For example, if a website has a complex UI, users would face difficulties to operate the site and eventually leave with dissatisfaction. Nevertheless, It is also suggested that the consumers respond positively to web pages which has a moderate range of complexity \cite{b3}.  With substantial complexities, users have to use greater cognitive resources and require a longer task completion time. So this is the reason why developers are constantly working on lessening cognitive load by adjusting the demands and diverse characteristics of users in the e-commerce industry \cite{b29}. From the web developers’ perspective, it is a top priority to keep consumers' interest by exhibiting a simple website interface \cite{b3} and providing a considerable experience so that users do not hesitate to visit that website again \cite{b2, b46}.
    
    In this paper, we briefly experimented with random individuals to describe illustrations of fixation on e-commerce websites by users. Therefore, we established a logical relationship between cognitive load and e-commerce aesthetic variables utilising contemporary technologies such as eye-tracking in our study quest. We scrutinised different websites and researched specific product pages, which is significant because it discloses an initial interaction with users before buying a product. This initial feeling is part and parcel of triggering users' assumptions for deciding whether they purchase from the site or not. Responsive web design comes with various elements such as graphics, content, figures, and hyperlinks, which influence purchasing behaviour. Also, website aesthetics strongly relates to webpage complexity \cite{b46}, and we anticipated that product price is one of the critical factors which plays an essential role to cause visual complexity, meaning different prices result in varying levels of complexity \cite{b47}. Apart from consumers’ perception, prices can be used to determine the amount of effort required to choose a product and the level of quality \cite{b6, b44}. This study is also devoted to figuring out the trigger of the price variable and how it influences a customer's purchasing habits.
    
\subsection{Statement of Problem}
    The complexity of the website defines the quality of its user interface and the mental impact on users' cognition or reaction. Besides, it all work simultaneously. K. Kidane and K.Sharma \cite{b32} further state that users' cultures and diverse perspectives affect consumers' complex behaviour in online shopping. Lots of effort need to comprehend visual perception and the underlying psychology behind making a site look good because developing it from the right visual angle is elemental. Research by Zhou et al. \cite{b33} states that a well-designed website is visually appealing for users. It also has a robust navigational function that allows potential consumers to browse the website more completely while avoiding obstructions. Finding relationships between website visual perception and users' cognitive load is essential to understand or enhance the visual perception of website efficacy. Similarly, finding the causes behind complex features and variables of websites seem necessary. Also, it is equally important to observe what makes users delay or offend choosing a required product while shopping online. In other words, how the price of a product impacts customer behaviour and what drives them to purchase specific goods. Most of the time, the orthodox way of developing a site is not enough to resolve the issue. The customer decision-making process is instant and unpredictable. We have worked with two methods to discover underlying reasons for users’ cognition and behaviour while shopping online. These methods distinguish user cognitive effort concerning website complexity on the e-commerce website. Besides, it would also assist us to find some connections of product price with users.
    
    %\vspace{1cm}

\subsection{Objective of Study}
    The main goal of our study is to find the relation between e-commerce product price and cognitive load. The study will discover the cognitive load effects on users’ purchasing behaviour in e-commerce websites to improve shopping experiences.
    
\subsection{Hypothesis}
    E-commerce websites mean the typical task of purchasing a product from an online retailer. Because of its convenience, e-commerce has become the preferred way of conducting shopping nowadays. Traditional websites with high complexity give plenty of information that need tremendous attention and time to observe and analyse \cite{b8}. Moreover, the perception of users and their cognitive load are naturally related so that the interaction between websites and humans happens immensely fast. Browsing interactions occur mainly in the subconscious part of our brain \cite{b12}, which result in quick response. If the cognitive load of e-commerce platforms are higher, it is needless to say that the interactivity will be laggard. Besides, consumer satisfaction and loyalty will be lost \cite{b11}. Also, a complex website will divert customers' attention and create more fixations \cite{b13}. In e-commerce research, CLT (Cognitive Load Theory) \cite{b14} is vastly considerable and influential, and we are motivated to investigate e-commerce websites to find out more variables and reasons for the varying cognitive load relation with the e-commerce website. As a result, when the website complexity rises, users' attention is more reasonable to expose task-irrelevant triggers, resulting in a higher fixation count, longer fixation duration, and longer task completion time \cite{b15, b8}. The proposed hypotheses are below:
    
    %\ref{itemtwo}
    %\label{itemtwo}
    
    \vspace{.1cm}
    \begin{enumerate}
        \item [\textbf{H1}.] The lower the website products’ price from the budget range, the higher the user’s fixation count and website complexity.
    \end{enumerate}
    \vspace{.2cm}
    \begin{enumerate}
        \item [\textbf{H2}.] The lower the website products’ price from the budget range, the longer the user’s task compilation time and the higher the website complexity.
    \end{enumerate}
    \vspace{.2cm}
    \begin{enumerate}
        \item [\textbf{H3}.] The lower the website products’ price from the budget range, the longer the user’s saccades and the higher the website complexity.
    \end{enumerate}
    \vspace{.2cm}
    \begin{enumerate}
        \item [\textbf{H4}.] The lower the website products’ price from the budget range, the longer the user’s fixation duration and the higher the website complexity.
    \end{enumerate}

\section{Literature Review}

\subsection{Background of the Study}

    Website visual perception has remained an active area of research for a long time. Plenty of tools and associate architects models have been developed and proposed for creating the adaptive user interface. Every day as technology advances, the interaction between humans and machines becomes more ubiquitous in our daily lives. Every advancement in the technology sector brings new challenges in our everyday lives if only the design of machines has not minimum useability \cite{b42}. Latest gadgets and applications allow immense possibilities and interactions between man and machine. HCD comes into practice to ensure this interactivity fluently. Human-centred design is the set of procedures to build a bridge between the people's capabilities and needs that fit appropriately with the system. Moreover, HCD stated that a subtle design begins with the utilisation of modern technologies and an understanding of human psychology \cite{b12}. It offers some vital principles prioritising human perception and focus compatibility of design by setting up a suitable approach. The failed development of HCD in websites create the complexity of the website. Besides, website complexity has a significant influence on user decision-making \cite{b43}.
    
    Suppose a user browse through a new website to perform a task. If the interaction of the newly visited website is not smooth enough to carry out the task with minimal use of cognitive load, then it will consider a complex website. The understanding of reciprocity between user and website depends on its low complexity. If a website's complexity is high, the user will find it unpleasant and difficult to interpret \cite{b2}. This complexity relies on many attributes where products price is one of them in e-commerce websites. Specifically, product price directly affects user purchasing decisions in online shopping \cite{b44}. To analyse and understand the user purchasing behaviour, CLT knowledge is required. In order to make a purchase or choose a product from an e-commerce website user’s decision-making process happens in the brain \cite{b14}. The decision-making process is influenced by several attributes \cite{b49}. To measure or manipulate these attributes, we used eye-tracking technology.
    
    By eye-tracking technology, a user's eye movement can be tracked and measured, which provide enough data to understand the influence of the user-specific behaviour \cite{b18}. Customers' gazing patterns can reveal valuable information about where they glance first while visiting a Web page, what portions of the page users give greater attention to, and how long they focus on those areas \cite{b46}.  A few eye-tracking events can extract the causes of cognitive load on users. In this study, we used two eye-tracking events: fixation and saccades. According to J. Zagermann et al., a user's focus time on a specific object or area is fixation, while the duration between two fixations is saccades \cite{b18}. By these two events, some study has shown users' interest and irritation while letting volunteers an online specific product purchase. Besides, the previous studies tried to find or hypothesise about the relation between users and websites \cite{b15}.
    
    The cognitive load and users' decision-making happens quite together while users shop online. To increase the user online shopping experience, improving e-commerce websites is significant. In our study, experiments are based on this core design philosophy. We focused on some aesthetics that caused impedance, making users woeful about their website browsing experience. We understand it is challenging to produce human-centric design, especially in e-commerce websites where good orientation is necessary so that consumers would surely revisit later.

\subsection{Related Work}

    Research on the nature of design characteristics and how to employ them have produced conceptual approaches and statistical findings. This section briefly explores those proposed reference architectures, models, adaptation techniques, and available tools, along with their limitations.
    
    It was mentioned earlier that low-visual-complexity graphic design has better usability and accessibility. So what exactly is the visual complexity? By definition, the degree of detail or delicacy present within a picture or a frame is referred to as visual complexity \cite{b30}. Tuch et al. \cite{b12} describe that a user's affective reaction during webpage perception is linked to Visual Complexity. Subjective feelings, behavioural expressions, and physiological responses are three levels of affective responding, and these differ as a function of Webpage Visual Complexity. In this way, a vital role in the adequate perception of Webpages is played by complexity. Similarly, Alexandre et al. \cite{b2} considered that visual complexity has an impact on valence and arousal judgments. They examined 36 sample website homepages to investigate the visual complexity for determining users' perception of psychological responses. Nevertheless, there is no clarity on what factors in website design cause visual complexity and how complexity is similar to several aspects of HCI, such as usability. However, an early study by Bruner and Kumar \cite{b20} revealed that the relationship between users and complex websites has many undeviating negative impressions from their three test subject sites, also mentioned in an earlier paper \cite{b22}. Conversely, positive reactions were also seen among website visitors in some aspects for visiting the complex sites, which happened indirectly. In other words, a website that has significant visual complexity can also be considered if some particular part or information of the webpage is worthwhile for users. Their outcome suggested that both positive and negative feedback would be possible having website complexity.
    
    Besides, another study focused on finding the relationship between website complexity and user's behaviour. Qiuzhen et al. \cite{b15} looked at the consolidated impact of website complexity and task complexity on users' actions and visual attention. The parameters such as task completion time, fixation count, and fixation duration distinguish task complexity (high or low). Although they maintained user activities with a website, such as clicking on the screen and eye movements using eye-tracking tools like us, they did not record individual perspectives while shopping online. In addition, Geissler et al. \cite{b3} examined the user's attention to the home page, attitude toward the home page, and purchase intent to measure the impact of home page complexity. They conducted two types of experiments, qualitative and pretest. In the qualitative process, they surveyed 30 web designers (both male and female) over a phone call to get an observation about web pages. On the other hand, 169 undergraduate students were the test subjects for the website evaluation's pretest process. As a result, their finding showed that users prefer intermediate complexity over less and higher complexity in an ordinary webpage. Also, minimal complexity led to users' maximum communication effectiveness. However, the users' emotional responses depend on the website's colour and contrast \cite{b28}, which was ignored in this experiment.
    
    Because online buying attracts people from various backgrounds and cultures, e-commerce enterprises are forced to acknowledge complex customer behaviour. As a result, studying the aspects that influence consumers' purchasing behaviour while shopping from e-commerce. Several related research explored the various aspects that influence customers' buying decisions. Lee and Kozar \cite{b10} suggested that a perceptive understanding of perceptions and purchase behaviour of consumers cannot be done without a proper recognition of website usability. This study explained website usability constructs and measurement tools and reviewed their nomological networks. On the other hand, Seo et al. \cite{b9} researched user emotional valence and perceived usability by exploring the relationships between users’ thoughts and emotional acknowledgements. However, the weak relationships between perceived usability and emotional arousal, the aforementioned finding for arousal may not have any implications. A study on usability satisfaction on user commitment was presented by Casalo et al. \cite{b27}. Still, there were gaps and limitations in existing techniques, such as the detailed way online users behaviour, privacy, security, or trust. Consequently, they proved that greater usability significantly impacts customer satisfaction.
    
    Some variables like colour schemes play a crucial role. Consumers are more likely to process information if there is a good contrast between the foreground and background colours. Jean and Panagiota \cite{b31} investigated how colours on websites affect user memory, emotion, and purchase intent. By examining the hue, brightness, and colour contribution of an e-commerce website, they presented a link between consumer memory, mood, and purchasing behaviour. Colour changes in web aesthetics also have a favourable impact on the user's ability to remember product information and purchasing intention. However, Lin et al. \cite{b28} found that users' emotional responses are negatively affected by the colour complexity of e-commerce websites. They have worked with some specific variables (figures, background colour and contrast) like we studied price as a critical aspect in this paper. Both them and we showed some variables directly responsible for causing visual complexity, which is antecedent to online shopping behaviours. Similarly, Bonnardel et al. \cite{b23} studied the aesthetic aspects of interaction with Websites. They experimented with some colours that were regarded as appealing or, on the opposite, were rejected by users and website designers. In particular, the homepages and websites were presented to participants in a single leading colour to perform a carefully controlled experiment, and the number of participants in each of the studies was reasonably small.
    
    One of the major findings from most research articles was that online shopping websites have extraneous cognitive load because of complicated structure and presentation. Ji et al. \cite{b21} came up with a method from the user’s cognitive response and behavioural information and constructed initial profiles from the 200 users dataset to create the cognitive-based model. As a result, they got positive outcomes with the customised UI/UX model compared to the traditional interface reviews for reducing cognitive load. Correspondingly, in their experiment, Peter et al. \cite{b29} showed how they measured cognitive workload using Nasa-TLX and dual-task methodologies. Their primary purpose was to find the user’s mental workload for different websites. They also investigated search preference and user satisfaction using many aspects of CLT \cite{b14}. Interestingly, Aladwani and Palviab \cite{b25} did empirical evaluation and verification by developing a multidimensional scale measuring user-perceived web quality without considering cognitive load. They showed that comprehensive testing and validation gained internal validity and used several variables that improved the external legality and generalizability to a larger population. On the contrary, Zhang et al. \cite{b26} proposed particular web environment aspects and sections that refer to a group of similar things. Nevertheless, the factors were the small representation size, and the specific information-seeking task within a particular domain were factors. Therefore, the effects of these different outcomes were not considered.
    
    Moreover, monitoring people's visual stimuli can assist HCI scholars to understand better the visual and display-based processing of information and the elements that influence system interface usability. Juni et al. \cite{b24} explained how to utilise the eye-tracking approach to assess consumer behaviours on web design or to discover customer interest in a particular product offered on an e-commerce website. Their work showed that it is hard to evaluate the user experience using pupil dilation because of the size difference of a human’s pupil. We did not face such consequences because the software we used to track eye movements was enough to calibrate the AOI. It runs several algorithms determining fixations concerning the relative position of the head and eyes \cite{b35}.
    
    Liqiong and Marshall \cite{b36} provided suggestions for showing and customising e-commerce webpages to improve user experience in practical applications. They identified various hyperlinks, images, and text as website decision variables that may impact webpage complexity and tested their efficiency by modifying various factors. They proposed that webpage complexity and order had substantial effects on consumer choice and a mediating effect on the client's buying attitude. Ultimately, they found no indication that the proposed recommendations had a moderating influence on website order preference, supporting our experiment with our test subjects.

\section{Methodology}

    \subsection{Data Collection}
        To gather raw data from the end-user, we used Google Forms and the Gazerecoarder \cite{b17}. Gazerecorder is a third-party software used for storing the details of every AOI (area of interest) based on the eye movements of the participants. AOI means the focus area of the user eyesight while browsing or looking at anything. The google form used for two purposes: collecting users personal information and every chosen product price. Then, we have selected the best 31 data set for the experiment out of 48 people responses. Besides, all participants had previous experience in online shopping. We conducted the experimentation under lab conditions to ensure its authenticity, quality and error-free. Similarly, all our participants had a support team to clear doubts about the survey.
        
        \begin{table}[htp]
            \caption{Population Sampling}
            \centering
            \begin{tabular}{|c|c|c|c|}
            \hline
            \multicolumn{3}{|c|}{\textbf{Characteristic}} & \multirow{2}{*}{\textbf{Total}} \\
            \cline{1-3}
                \textbf{Age Group} & \textbf{Male} & \textbf{Female} &\\
                \hline
                    18-25 & 7 & 3 & 10 \\
                \hline
                    26-33 & 8 & 6 & 14  \\
                \hline
                    34-41 & 4 & 2 & 6  \\
                \hline
                    42-50 & 1 & 0 & 1  \\
                \hline
                    \textbf{In total} & 20 & 11 & 31  \\
                \hline
            \end{tabular}
            \label{tab:fullwidth}
        \end{table}

        In the first segment of our data collection, we collected some basic personal details about our volunteers who participated in the experiment. We carefully considered the device specification (computer screen size and webcam type) to perform the study under a similar environment for everyone. Afterwards, we redirect the survey-taker to the gazerecorder to track their eye movements.
        
        \begin{figure}[htp]
            \centerline{\includegraphics[width=0.4\textwidth, height=8cm]{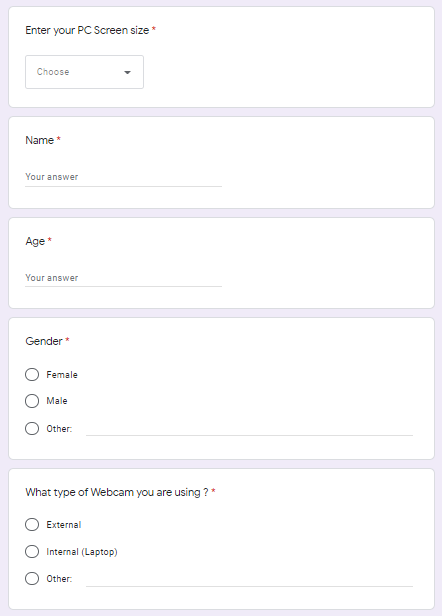}}
            \caption{Survey Form}
        \end{figure}
        
        After reaching the gazerecorder website, the web application first scanned the person face through a webcam where it detected the eyeball by the movements of a redpoint. When it successfully encountered the eye-rolling gesture, it took the user to the final study page, where actual data collection occurred.
        
        \begin{figure}[htp]
            \centerline{\includegraphics[width=0.5\textwidth,
            height=5cm]{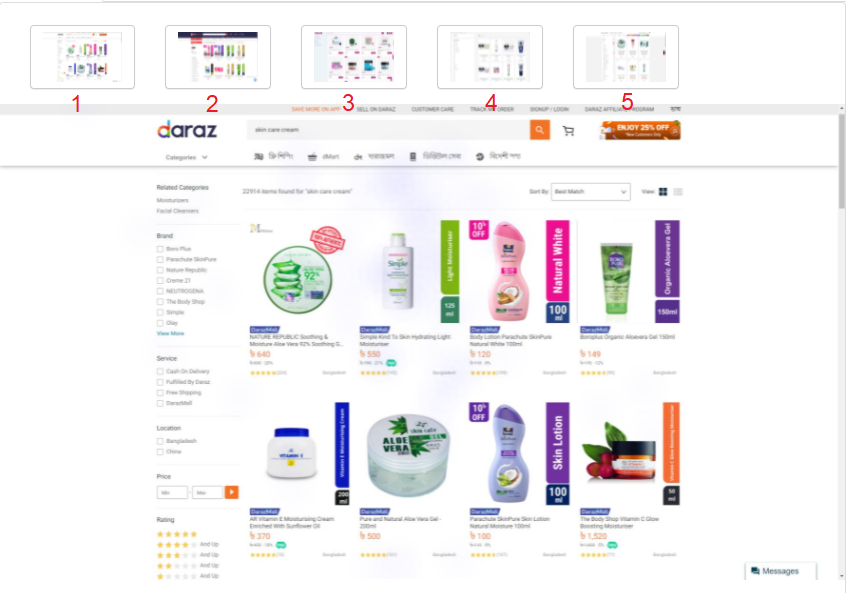}}
            \caption{Study Webpages}
        \end{figure}
        
        We have chosen five different e-commerce websites and set up five individual static pages by taking screenshots like Alexandre et al. \cite{b2} and Liqiong et al. \cite{b36} did in their stimuli process. Those websites were displayed in the gazerecoder slideshow in front of volunteers to perform the study. We showcased some skincare cosmetics on those web pages where the products were popular in the region. Moreover, we instructed users to select one product from each web page. The duration of choosing the product from a single page was 45 seconds, where we did not give any constraint about selecting any products of their wish except budget. As a result, the participants had chosen one product from each page with complete freedom under the given time. After 45 seconds, the page changes automatically and appears on the following website. Besides, every participant was about 75 cm (closely 2.5 feet) apart or inside from the computer screen. The total time of the experiment required for each volunteer was about 5-7 minutes. To perceive the AOI of the study and the heat map of eye-tracking data was generated automatically in the gazerecorder’s system. After each study completion, users rated their experience on a scale of 1 to 5 and put down some suggestions. Around 60\% of them rated the survey 5.
    
    %\vspace{0.5cm}  
    
    \subsection{Data Processing}
        
        For better understanding, Table~\ref{tab:table-Work-flow}  illustrate the workflow of the data collection and calculation process. In this stage, the authors' primary objective was to go through the raw data and analysis step by step until they reached new findings of our result.
        
        \begin{table}[h!]
        \caption{Data Collection Work-flow}
            \centering
            \begin{tabular}{|c|}
            \hline
                \textbf{Data Collection} \\
                \hline
                    Data Collected from users’ survey \\
                \hline
            \end{tabular}
            \label{tab:table-Work-flow}
        \end{table}
        
    % \vspace{-0.25cm}  
     
        \begin{table}[htp]
            \centering
            \begin{tabular}{|c|}
            \hline
                \textbf{Data Pre Processing} \\
                \hline
                    Classifying the Websites \\
                \hline
                    Inspection and point out on the heatmaps \\
                \hline
                    Counted multiple dwell’s time and measured \tabularnewline saccades from perceived data \\
    
                \hline
                    Calculated task completion time, total fixation duration \\
                \hline
                    Calculate the Standard Deviation of fixation  \\
                \hline
            \end{tabular}
        \end{table}
    
% \vspace{-0.25cm}
    
        \begin{table}[htp]
            \centering
            \begin{tabular}{|c|}
            \hline
                \textbf{Data Analysis with respective methods} \\
                \hline
                    Method 1: Fixation \\
                \hline
                    Method 2: Saccades \\
                \hline
                    Method 3: Statistical Analysis \\
                \hline
            \end{tabular}
        \end{table}

        \subsubsection{Classification of Websites}
        
            We classified the five websites using the decision tree before added in the survey. The decision tree is used in statistics to construct data classification or regression.
            
            \begin{figure}[htp]
            \centerline{\includegraphics[width=0.5\textwidth,
            height=5cm]{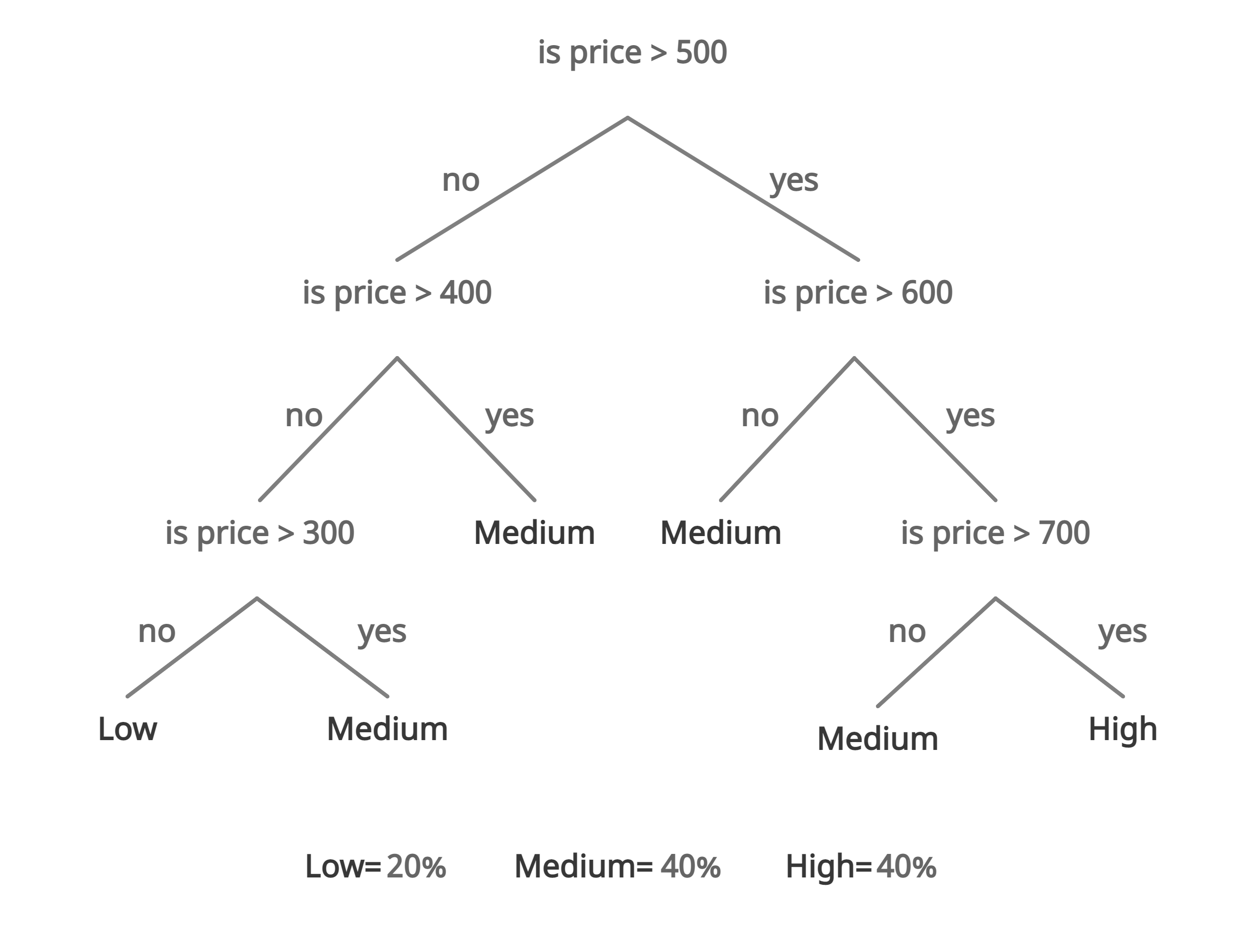}}
            \caption{The decision tree of the website's classification}
            \label{fig:figure-Classification}
        \end{figure}
            
            Figure~\ref{fig:figure-Classification} shows the three classifications of the websites, which are respectively low, medium, and high. The decision process is created based on product price and user budget. Besides, most of the products are well known and popular where we surveyed, and the websites are also in the leading position in the region.

            \begin{table}[htp]
            \caption{Website Classification}
                \centering
                \begin{tabular}{|c|c|}
                \hline
                    \textbf{Name} & \textbf{Classification}\\
                    \hline
                        BanglaShoppers & Low \\
                    \hline
                        Daraz & Medium \\
                    \hline
                        PriyoShop & Medium \\
                    \hline
                        Shajgoj & High \\
                    \hline
                        Paikaree & High \\
                    \hline
                \end{tabular}
                \label{tab:table-Classification}
            \end{table}
    
    %\vspace{0.3cm}   
    
            However, Table~\ref{tab:table-Classification}. data is qualitative data, not numerical. As a result, we must first transform the qualitative data into a numeric form to conduct our experiment. So we have assigned numerical values of the website classification instead of categorical data. We used scaling websites from 1 to 3 where ascending order represents low to high in Table~\ref{tab:table-Numeric-Classification}
    
    %\vspace{0.2cm}
    
            \begin{table}[h!]
            \caption{Website Numeric Classification}
                \centering
                \begin{tabular}{|c|c|}
                \hline
                    \textbf{Name} & \textbf{Value}\\
                    \hline
                        Law & 1 \\
                    \hline
                        Medium & 2 \\
                    \hline
                        Medium & 2 \\
                    \hline
                        High & 3 \\
                    \hline
                        High & 3 \\
                    \hline
                \end{tabular}
                \label{tab:table-Numeric-Classification}
            \end{table}
        
        \subsubsection{Heatmap Inspection}
    
        A heatmap illustrated users' eye movement and continued AOI with great details. Before jumping into the actual calculation, we needed to observe heatmaps deliberately to find out users' actions and intentions. We primarily tried to find out the pattern of users activity and endeavoured to seek the gap between action and preference. In other words, heatmap analysis allows us to compare the survey data with the actual visual interest of volunteers for choosing the product. Here, heatmap generation is a tracked footprint of eye movements and interest.
    
    %\vspace{0.5cm}
    
            \begin{figure}[htp]
                \centerline{\includegraphics[width=0.5\textwidth, height=5cm]{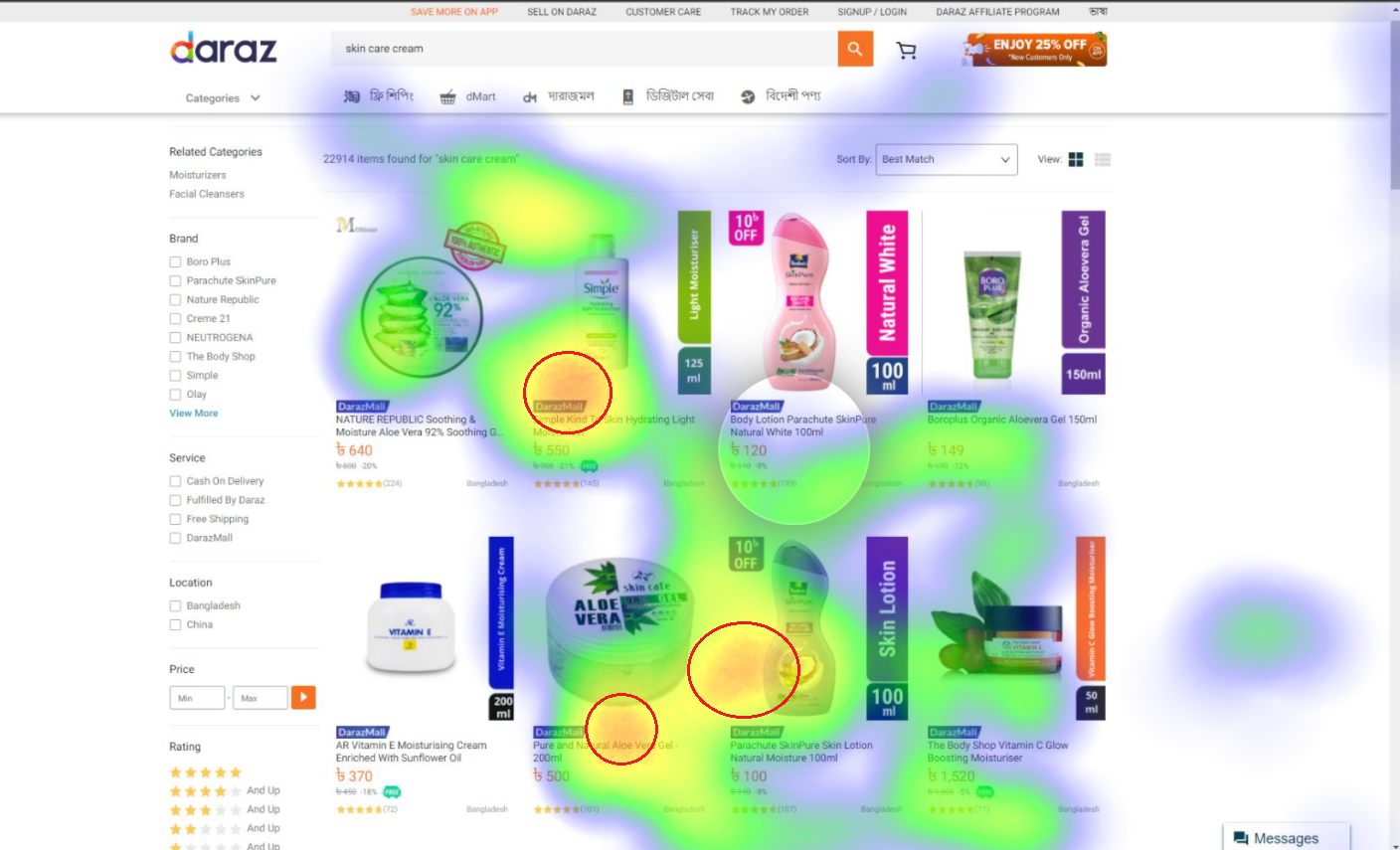}}
                \caption{Detailed analysis of a heatmap}
                \label{fig-heatmap}
            \end{figure}
            
    %\vspace{0.2cm}    
    
        Fig.~\ref{fig-heatmap} illustration shows the recorded eye movement of a volunteer while buying a product. The contrast of colours described the duration of the time difference between one to another AOI. In other words, all footprints of the users represent the visual interest or interested product. The more profound contrasted heatmap refers to the more time a user viewed that particular spot. We marked down the most heated area with red circles to display the significance of AOI. These helped us understand that visual interest anticipates AOI count so that we could reach out to the actual outcome of the study. Besides, three red circles represent the maximum AOIs of the heatmap. Hence we can say that the user concentrated carefully on those products. Although these three red spots are pretty similar from an observer point of view, we settle down to three possible outcomes or AOI counts by assuming one of these might be the chosen product. Therefore, the giant circle on the picture is the actual selected product by the test-taker. We had known that from the survey data, we would also observe it from the AOI covered area on the calibration picture. In addition, more heated footprints convey the higher complexity of a website \cite{b15}. That means when a website grabs way more attention from a user and generates extensive impressions in the heatmap; it indicates the higher complexity of the website \cite{b16}. Therefore, a larger AOI count also refers to the higher complexity.

        \subsubsection{Analyse the dataset for finding-out basic attributes}
            
            We classified and distinguished relevant and irrelevant data from eye-tracking software and named those data under different attributes. We also trimmed the user spent on the task data to ensure the accurate time. To categorise the raw data, we used three variables. They are sequentially first view, last view and dwell time of a single AOIs as shown in Table~\ref{tab:table-Basic-attributes}.
     
            \begin{table}[htp] 
            \caption{Basic  attributes}
            \centering 
            \begin{tabular}{|M{0.7cm}|M{0.7cm}|M{1.2cm}|M{0.8cm}|M{0.7cm}|M{1.2cm}|}
            \hline
                \textbf{First view(s)} & \textbf{Last view(s)} & \textbf{1st Dwell Time(s)} & \textbf{Second view(s)} & \textbf{Last view(s)} & \textbf{2nd Dwell Time(s)} \\
                \hline
                    0.57 & 1.6 & 1.03 &	0 &	0 &	0 \\
                \hline
                    0.86 & 2.8 & 1.94 & 0 & 0 & 0 \\
                    \hline
                    0.76&	1.29&	0.53&	6.1	&6.6&	0.5\\
                \hline
                    1.54&	2.52&	0.98&	0&	0&	0\\
                \hline
                    1.16&	1.91&	0.75&	8.6&	9.9	&1.3\\
                \hline
                    0.78&	1.27&	0.49&	0&	0&	0\\
                \hline
                    0.58&	3.28&	2.7	&10.53&	11.31&	0.78\\
                \hline
                    4.37&	5.07&	0.7&	0&	0&	0\\
                \hline
            \end{tabular}
            \label{tab:table-Basic-attributes}
        \end{table}
        
        \subsubsection{Calculating specific attributes from basic attributes}
        
            After the data collection, the data have been checked and verified by matching the chosen product on both surveys (Google Forms and Gazerecorder) outputs. There are Five significant attributes which are saccades, task completion time (TCT), Total Fixation Duration (TFD), and Total Fixation (TF), needed to find out for calculating the methods. Those attributes have some pre-attributes to estimate first from raw data.
            
            \renewcommand{\labelenumii}{\arabic{enumi}.\arabic{enumii}}
            \renewcommand{\labelenumiii}{\arabic{enumi}.\arabic{enumii}.\arabic{enumiii}}
            \renewcommand{\labelenumiv}{\arabic{enumi}.\arabic{enumii}.\arabic{enumiii}.\arabic{enumiv}}
            \begin{enumerate}
                \item Fixation
                    \begin{enumerate}
                        \item Fast view of AOI
                        \item Last view of AOI
                    \end{enumerate}
                \item Total Fixation
                \item Average Task Completion Time
                    \begin{enumerate}
                        \item The first view of the very first fixation
                        \item The last view of the last fixation
                    \end{enumerate}
                \item Total Fixation Duration
            \begin{enumerate}
                \item Sum of Multiple same AOI’s dwell time
            \end{enumerate}
            \item Single Saccades 
                \begin{enumerate}
                    \item Last view of a Fixation
                    \item First view of the next Fixation
                \end{enumerate}
            \end{enumerate}
    
            \textbf{Here, TFD + Saccades = TCT}
            
    \subsection{Eye tracking and task measurement}
    
        Eye gestures are influenced by perceived world features and human mental processes \cite{b48}. Using eye movement tracking tools to investigate persons' web activities provides a deeper understanding of purchasing behaviour and cognition \cite{b46}. We used the most popular two different eye-tracking events to calculate the user's cognitive load effect while shopping online. These two events are Fixations and Saccades, where both have different characteristics and attributes. On the other hand, to measure user tasks or a variable to measure how long it takes to complete work for a user is used as a task compilation time. 
    
    %\vspace{0.3cm}
    
    \subsubsection{Fixation}
        
        An eye-tracking event is when the eye remains focused on a selected area for a long time. It is known as Fixation or AOI, and it can last anywhere from 200-300 milliseconds to several seconds. A Fixation has two attributes which are Fixation count and Fixation duration. Fixation count means the total AOI in a single task compilation time. On the other side, the difference between the First View and Second View of an AOI is called fixation duration or dwell time. The cognitive load influenced fixation duration and Fixation count; the higher the cognitive load on the user, the greater the Fixation count and Fixation duration \cite{b18}.
            
    %\vspace{0.3cm}
    
        \subsubsection{Saccades}
        
            Saccade refers to a voluntary movement of eyes from one Fixation to the next. It is the quickest movement the human body can do, and it takes between 30-80 milliseconds to complete. Like Fixation, Saccade also has two attributes. Those are Saccade length and Saccade velocity. The duration of an eye movement from one Fixation to the next is called a Saccade length, but in this study, it is a saccade. The second attribute of Saccade is called saccade velocity, which explains how fast the eye move. The technology used in this research were not enough to calculate the saccade velocity. In contrast to cognition, the higher the cognitive load on the user, the longer the saccades \cite{b18}.
            
        \subsubsection{TCT}
        
            The duration between the First view of the very first Fixation and the last view of the Last Fixation is the TCT. TCT means the user’s total spending time to choose a product or Total Compilation Time.
            
    \subsection{Correlation coefficient}
    
        Pearson's r analysis is one of the most extensively used and reported statistical procedures in evaluating biomedical research and scientific data. It's frequently helpful to see a connection between two independent variables. A study found that its result calculates from -1.0 to 1.0. A positive value means that an increase in one variable will rise in the other. In contrast, a negative correlation denotes an inverse relationship where one variable increases and the other fall. On the other hand, a value of 0 correlation indicates no link between the two variables' or data set \cite{b19}.

\section{Results and Analysis}
    
    As we mentioned, we collected our dataset from the survey. Here we are trying to find out the reaction between users task compilation time on the webpage and website products’ pricing; considering website complexity.
    
    In methodology, there is mention of data processing and heatmap inspection. Here in the result and analysis part, different kinds of calculation will be discussed, which we apply to our dataset and find some results. Regarding those results, here we will try to explain the results.

    \subsection{Hypotheses testing}

    \begin{table}[htp]
            \caption{Results of Single-factor ANOVA}
        \begin{tabular}{|c|cccc|}
        \hline
        \multirow{2}{*}{\textbf{\begin{tabular}[c]{@{}c@{}}Independent \\ Variable\end{tabular}}} & \multicolumn{4}{c|}{\textbf{Dependent Variables}}  \\
        \cline{2-5} 
        & \multicolumn{1}{c|}{\textbf{\begin{tabular}[c]{@{}c@{}}Fixation\\ count\end{tabular}}} & \multicolumn{1}{c|}{\textbf{TCT}} & \multicolumn{1}{c|}{\textbf{Saccades}} & \textbf{\begin{tabular}[c]{@{}c@{}}Fixation\\ duration\end{tabular}} \\ 
        \hline
        \multirow{3}{*}{\textbf{\begin{tabular}[c]{@{}c@{}}Website \\ Complexity\end{tabular}}}   & \multicolumn{1}{c|}{F= 123.85} & \multicolumn{1}{c|}{F = 115.23}   & \multicolumn{1}{c|}{F = 6.49} & F = 9.90 \\ \cline{2-5} 
         & \multicolumn{1}{c|}{F crit = 6} & \multicolumn{1}{c|}{F crit = 5.9} & \multicolumn{1}{c|}{F crit = 6} & F crit = 5.98 \\ \cline{2-5} 
        & \multicolumn{1}{c|}{P = 0.000} & \multicolumn{1}{c|}{P = 0.000} & \multicolumn{1}{c|}{P = 0.043} & P = 0.000 \\ 
        \hline
        \end{tabular}
        \label{tab:table-single-factor}
    \end{table}

    Table~\ref{tab:table-single-factor} shows every F $>$ F crit and p $<$ 0.05 so that hypotheses are supported. However, for data directions and differences, we used the t-test to test each pair of means of cognitive variables.

    To understand the user's purchasing behaviour with the price of the website product, we have used a hypothesis test (T-test). The fixation count on the website with low categorization is significantly higher than those on the website with medium classification, according to the findings of a t-test. The mean difference between Low and medium is 2, and their p-value is 0.011. Similarly, for medium and high mean difference = 1, p = 0.031. And for Low and High = 3, p =  0.03. H1 is accepted (p<0.05). For task compilation time, mean difference (low and medium) = 2.695, p = 0.000. Likewise, for medium and high mean difference = 0.73, p = 0.000. And for Low and High = 3.425, p =  0.000. H2 is accepted. The saccades with the low classified website are significantly higher than medium. Besides, Mean difference of (low and medium) = 0.455, p = 0.053. Again, for medium and high mean difference = 0.85, p = 0.01. And for Low and High = 0.395, p =  0.039. H3 is partially accepted. For fixation duration, the mean difference between low and medium =  3.422, p = 0.01. More, for medium and high mean difference = 0.21, p = 0.04. And for Low and High = 3.34, p =  0.000 H4 is accepted. Thus we can see that the effects of cognitive load on user purchase behaviours are significant (p $<$ 0.05).

    \begin{table}[htp]
    \caption{The t-test of users behaviours variables}
        \centering
        \begin{tabular}{|c|c|c|c|c|c|} 
        \hline
        \multirow{2}{*}{\begin{tabular}[c]{@{}c@{}}\textbf{Independent }\\\textbf{Variables }\end{tabular}} & \multicolumn{5}{c|}{\textbf{Dependent Variables }} \\ 
        \cline{2-6}
         &  & \begin{tabular}[c]{@{}c@{}}\textbf{Fixation}\\\textbf{Count}\end{tabular} & \textbf{TCT} & \textbf{Saccades} & \begin{tabular}[c]{@{}c@{}}\textbf{Fixation}\\\textbf{Duration}\end{tabular} \\ 
        \hline
        \multirow{2}{*}{\begin{tabular}[c]{@{}c@{}}\textbf{Low }\\\textbf{vs}\\\textbf{~Medium }\end{tabular}} & \begin{tabular}[c]{@{}c@{}}T \\stat\end{tabular} & 2.414 & 6.055 & -1.661 & 2.436 \\ 
        \cline{2-6}
         & \begin{tabular}[c]{@{}c@{}}T \\critical\end{tabular} & 1.699 & 1.705 & 1.699 & 1.699 \\ 
        \hline
        \multirow{2}{*}{\begin{tabular}[c]{@{}c@{}}\textbf{Medium}\\\textbf{vs}\\\textbf{High }\end{tabular}} & \begin{tabular}[c]{@{}c@{}}T \\stat\end{tabular} & 1.954 & 7.792 & 2.328 & 1.706 \\ 
        \cline{2-6}
         & \begin{tabular}[c]{@{}c@{}}T \\critical\end{tabular} & 1.703 & 1.703 & 1.701 & 1.703 \\ 
        \hline
        \multirow{2}{*}{\begin{tabular}[c]{@{}c@{}}\textbf{Low}\\\textbf{vs}\\\textbf{High }\end{tabular}} & \begin{tabular}[c]{@{}c@{}}T \\stat\end{tabular} & 4.048 & 6.84 & 1.824 & 4.525 \\ 
        \cline{2-6}
         & \begin{tabular}[c]{@{}c@{}}T \\critical\end{tabular} & 1.701 & 1.703 & 1.701 & 1.699 \\
        \hline
        \end{tabular}   
    \label{tab:table-t-test-user-behaviours}
    \end{table}
    
    Table~\ref{tab:table-t-test-user-behaviours} shows the t-stat and t Critical one-tail values. Based on this data, H1, H2 and H4 are supported while H3 is partly supported. Because the T stat value is greater than T critical for Low vs Medium.

    \subsection{Relation of users’ purchase behaviour with product price}
    
    \begin{table}[htp]
        \caption{Descriptive statistics}
    \centering
        \begin{tabular}{|c|c|c|c|c|} 
        \hline
        \multirow{2}{*}{\textbf{Dependent Variables }} & \multicolumn{4}{c|}{\textbf{Independent Variables }} \\ 
        \cline{2-5}
         &  & \textbf{Low} & \textbf{Medium} & \textbf{High} \\ 
        \hline
        \multirow{2}{*}{\textbf{Fixation Count }} & Mean & 9 & 7 & 6 \\ 
        \cline{2-5}
         & SD & 3.69 & 3 & 2.95 \\ 
        \hline
        \multirow{2}{*}{\textbf{TCT }} & Mean & 15.86 & 13.17 & 12.435 \\ 
        \cline{2-5}
         & SD & 9.16 & 6.62 & 5.6 \\ 
        \hline
        \multirow{2}{*}{\textbf{Saccades }} & Mean & 4.86 & 5.32 & 4.47 \\ 
        \cline{2-5}
         & SD & 3.51 & 4.41 & 3.38 \\ 
        \hline
        \multirow{2}{*}{\textbf{Fixation Duration }} & Mean & 11.08 & 7.95 & 7.74 \\ 
        \cline{2-5}
         & SD & 6.62 & 3.77 & 3.18 \\
        \hline
        \end{tabular}
        \label{tab:table-descriptive-statistics}
    \end{table}
    
    Fixation count is related to website complexity. The higher the number, the higher the complexity \cite{b15}. Table~\ref{tab:table-descriptive-statistics} shows that the relation between website price and fixation count is inversely proportional. Website complexity increases if the cognitive load of user’s increases. The complexity of a website has an impact on users fixation duration and TCT. Table~\ref{tab:table-descriptive-statistics} shows if the website price increase, then the fixation duration and TCT decrease. As a result, the user takes a long time to complete the task \cite{b15}. A single saccade is just a few milliseconds long events, and we have already explained the saccades in the methodology. However, saccades didn’t follow the same pattern as fixation count, TCT and fixation duration but its low and medium classified are both greater than high classified websites.
    
    \subsection{Correlation coefficient of the product price with the dataset}
    
        We calculated the correlation coefficient of the data sets to see the connection between cognitive variables and the product's price of e-commerce websites.
        
    \begin{table}[htp]
    \caption{Pearson correlation coefficient}
    \centering
        \begin{tabular}{|c|c|c|c|c|} 
        \hline
         & \begin{tabular}[c]{@{}c@{}}\textbf{Fixation}\\\textbf{Count}\end{tabular} & \textbf{TCT} & \textbf{Saccades} & \begin{tabular}[c]{@{}c@{}}\textbf{Fixation}\\\textbf{Duration}\end{tabular} \\ 
        \hline
        \textbf{Coefficient r} & -0.6767 & -0.6281 & -0.1573 & -0.6975 \\
        \hline
        \end{tabular}
        \label{tab:table-correlation-coefficient}
    \end{table}
    
    According to Table~\ref{tab:table-correlation-coefficient}, the correlation coefficient of fixation count, TCT and fixation duration are sequentially -0.6767, -0.6281 and -0.6975. It denotes a link of products’ price and users cognitive load in a website. Besides, the correlation coefficients are negative; that means the strength of the relationship between these relative movements of two variables is inversely proportional. However, the correlation coefficient of saccades is -0.1573 indicating the relation between product price and saccades are very poor and inversely related.

\section{Discussion}
\begin{figure}[htp]
        \centerline{\includegraphics[width=0.5\textwidth, height=6cm]{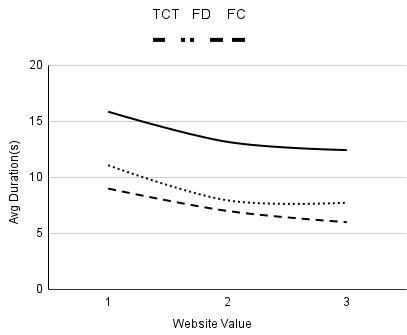}}
        \caption{The individual average of TCT, FD and FC}
        \label{fig-TCT and FD}
    \end{figure}

    Based on the findings of this research, products’ prices has an essential role in e-commerce sites. When users perform primary activities \cite{b29} like choosing a product on websites with a lower price range, task completion time is longer on a webpage. In other words, the lower the product price from the customer budget, the longer it takes time to decide on a particular product. It increases the time of users decision-making process, which is why they spend more time than usual online shopping. It also creates distractions while shopping from e-commerce sites. Correspondingly, the lower the product price, the higher the user fixation count at the screen to settle down for choosing a product. It also distracts the user from shopping. Likewise, the total time of fixation duration increases. Fig.~\ref{fig-TCT and FD} shows the average task compilation time, fixation count and fixation duration in the same pattern. The higher the website complexity, the longer the duration of TCT, FD and the higher the fixation count. It justifies our H1, H2 and H4 hypotheses that low price products grab more attention than typical from the user than medium and high price range products.
    
    %\vspace{1cm}

\begin{figure}[htp]
        \centerline{\includegraphics[width=0.5\textwidth, height=6cm]{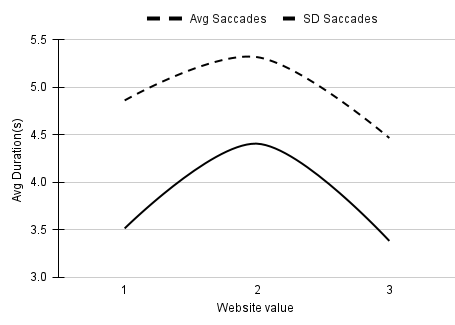}}
        \caption{The average and standard deviation of Saccades}
        \label{fig-Fixation and Saccades}
    \end{figure}

    However, saccades did not follow this pattern for the low classification. The standard deviation of saccades and average of saccades follow the same rules and slightly contradict the H3, as Fig.~\ref{fig-Fixation and Saccades} shows. It shows a longer duration in medium price ranged websites than the low classified websites. We know that cognitive load refers to the quantity of information that STM can process at each moment. Cognitive load theorist Sweller \cite{b14} suggests that working memory has a limited capacity. So, It is essential to prevent overloading it with extraneous tasks that do not directly contribute to knowledge or the decision-making process.
    
    \subsection{Future Work}
    
        As this is our undergraduate thesis, we narrowed down the goal due to the limited period. Hence, we did not work on any model that can help compare or find the exact issues in the website. An ideal model for e-commerce especially considering cognitive load, would do a great job. Nowadays, website developers always follow a few same rules to build a website. From that perspective, a User-interface (UI) model for an e-commerce site can effectively change the current viewpoint for anyone who is trying to build or own a new one.
        
        Here are some limitations: we can not explain a particular participant’s intention or action. It was impossible to define the exact reason for a person’s AOI based on eye-tracking data without psychological queries. Moreover, a person’s intention and action are complex, with plenty of inconsistent concern about what makes a person do a specific task. Consumers have a variety of incentives while shopping online. It depends on how they engage with the aesthetics of a website which involves a wide range of activities [34]. As a result, we estimated the visual perception of the website instead of individual AOI motivations. In the future, we want to find out the reasons behind every gaze made on the screen by a person.
        
        This study was conducted in the web environment, especially using computers. Our study empirically discovers some key variables which influence customers in their purchasing. However, the shopping trend is continuously evolving, so the user adaptations. Nowadays, a shopping cart is available through mobile applications, which is more convenient than browsing on computers. Our extended version of this study will carefully consider the mobile domain. In the future study, we will attempt to define characteristics in a mobile context about users' perceptions and e-commerce culture.
        
        Technology is progressing at a fast pace than we could have anticipated and what we are currently using is very different from those we used ten years ago. As a result of technological advancement, using advanced technology in the same research will result in less erroneous and more accurate data, which will lead to nearly perfect results and new findings.
        
        There are numerous variables while users browse a website related to the users working memory and cognitive load. The environment is one of the variables that influence users most. During website browsing, a buyer always tries to make the final decision. In the meantime, any surroundings can drive that happen. To find out exactly what component is significantly involved with decision-making requires more time and knowledge to get an endpoint. In addition, we have a plan to gear up our study by observing many other external responses like heart rate, expression detection and facial temperature detection while conducting the study between human and computer interaction.

\section{Conclusion}
    Research on users behaviour toward online shopping and the relation between cognitive load and website complexity has provided more knowledge of the nature of e-commerce websites and customers. Previous findings suggest that the website complexity effect influence users cognition and decision. However, there is a lack of clear relations and familiarity between customers and e-commerce because a single website contains many factors related to consumers' decision-making process. Our research focuses on the relationship between product pricing and the impact of cognitive load on a user's purchasing behaviour while shopping on an e-commerce website. Besides, product choosing happens on the user's short-term memory, which means product choosing happens faster. However, the process is fragile due to decision efficiency and high website complexity that affects their purchase behaviour. For instance, many items with the exact pricing require customers to take more time to choose. People usually do online shopping to keep it hassle-free and save time. Therefore, we specifically work with this problem to find the interrelation so that business analysts and product managers can use our study results as guidance. Nonetheless, it also gives the correlation between user behaviour and website complexity which will help the web developers better understand how product prices drive or potential influence customers. Moreover, the methods of this study show how to overcome the typical system analysis from complaints or feedback of customers to improve the e-commerce services. This research also demonstrates how to leverage the physiological data of customers without interrogating them. Although we are limited to vast resources and environments, we powered the experiment using third-party software and intricately analysed all of the datasets ourselves. In upcoming times, we aim to build our eye-tracking system to automatically get the desired data from the scanned outcome and plot graphs \& charts to see the instant result of the data. Further research is necessary to answer human perception measuring external responses concerning web variables to reduce cognitive load.

\vspace{12pt}

\end{document}